\documentclass[twocolumn]{aastex62}

\newcommand{\eg}{e.g.,\ }

\newcommand{\Msun}{$M_{\odot}$}

\newcommand{\kms}{km~s$^{-1}$}

\newcommand{\SiI}{Si~{\sc i}}

\newcommand{\SI}{S~{\sc i}}

\newcommand{\CaII}{Ca~{\sc ii}}
\newcommand{\CaI}{Ca~{\sc i}}

\newcommand{\FeII}{Fe~{\sc ii}}

\newcommand{\CoII}{Co~{\sc ii}}

\newcommand{\NiII}{Ni~{\sc ii}}

\newcommand{\Nifs}{$^{56}$Ni}

\newcommand{\KE}{$E_{\rm kin}$}

\newcommand{\sBV}{s$_{BV}$}

\newcommand{\ab}{$\sim$}
\newcommand{\mic}{$\mu$m}

\newcommand{\DmsV}{\Delta{\rm m_{15,s}(V)}}
\newcommand{\DmsB}{\Delta{\rm m_{15,s}(B)}}
\newcommand{\ved}{$v_{edge}$}

\graphicspath{{./}{figures/}}

\accepted{\today}
\submitjournal{ApJ}

\shorttitle{\ved\ probing \Nifs}
\shortauthors{Ashall et al.}
\begin{document}

\title{A Physical Basis for the H-band Blue-edge Velocity and Light-Curve Shape Correlation  in Context of Type Ia Supernova Explosion Physics}

\correspondingauthor{Chris Ashall}
\email{Chris.Ashall@gmail.com}

\author{C. Ashall}
\affil{Department of Physics, Florida State University, Tallahassee, FL 32306, USA}

\author{P. Hoeflich}
\affiliation{Department of Physics, Florida State University, Tallahassee, FL 32306, USA}

\author{E. Y. Hsiao}
\affiliation{Department of Physics, Florida State University, Tallahassee, FL 32306, USA}

\author{M. M. Phillips}
\affil{Department of Physics, Florida State University, Tallahassee, FL 32306, USA}
\affiliation{Carnegie Observatories, Las Campanas Observatory, 601 Casilla, La Serena, Chile }

\author{M. Stritzinger}
\affiliation{Department of Physics and Astronomy, Aarhus University, 
Ny Munkegade 120, DK-8000 Aarhus C, Denmark }

\author{E. Baron}
\affiliation{Department of Physics and Astronomy, Aarhus University, 
Ny Munkegade 120, DK-8000 Aarhus C, Denmark }
\affiliation{Homer L. Dodge Department of Physics and Astronomy, University of Oklahoma, 440 W. Brooks, Rm 100, Norman, OK 73019-2061, USA}
\affiliation{Hamburger Sternwarte, Gojenbergsweg 112, D-21029 Hamburg, Germany}

\author{A. L. Piro}
\affiliation{Observatories of the Carnegie Institution for Science, 813 Santa Barbara St., Pasadena, CA 91101, USA }

\author{C. Burns}
\affiliation{Observatories of the Carnegie Institution for Science, 813 Santa Barbara St., Pasadena, CA 91101, USA }

\author{C. Contreras}
\affiliation{Carnegie Observatories, Las Campanas Observatory, 601 Casilla, La Serena, Chile}

\author{S. Davis}
\affiliation{Department of Physics, Florida State University, Tallahassee, FL 32306, USA}

\author{L. Galbany}
\affiliation{PITT PACC, Department of Physics and Astronomy, University of Pittsburgh, Pittsburgh, PA 15260, USA}

\author{S. Holmbo}
\affiliation{Department of Physics and Astronomy, Aarhus University, 
Ny Munkegade 120, DK-8000 Aarhus C, Denmark }

\author{R. P. Kirshner}
\affiliation{Gordon and Betty Moore Foundation, 1661 Page Mill Road, Palo Alto, CA 94304 }
\affiliation{Harvard-Smithsonian Center for Astrophysics, 60 Garden Street, Cambridge, MA 02138}

\author{K. Krisciunas }
\affiliation{George P. and Cynthia Woods Mitchell Institute for Fundamental Physics \& Astronomy, Texas A\&M University, Department of Physics, 4242 TAMU, College Station, TX 77843}

\author{G.~H.~Marion}
\affiliation{University of Texas at Austin, 1 University Station C1400, Austin, TX, 78712-0259, USA}

\author{N. Morrell}
\affiliation{Carnegie Observatories, Las Campanas Observatory, 601 Casilla, La Serena, Chile }

\author{D. J. Sand}
\affil{Department of Astronomy/Steward Observatory, 933 North Cherry Avenue, Rm. N204, Tucson, AZ 85721-0065, USA}

\author{M. Shahbandeh}
\affiliation{Department of Physics, Florida State University, Tallahassee, FL 32306, USA}

\author{N. B. Suntzeff}
\affiliation{George P. and Cynthia Woods Mitchell Institute for Fundamental Physics \& Astronomy, Texas A\&M University, Department of Physics, 4242 TAMU, College Station, TX 77843}

\author{F. Taddia}
\affiliation{Department of Physics and Astronomy, Aarhus University, 
Ny Munkegade 120, DK-8000 Aarhus C, Denmark }
%\affiliation{ The Oskar Klein Centre, Department of Astronomy, Stockholm University, AlbaNova, 106 91 Stockholm, Sweden}

\begin{abstract}
Our recent work demonstrates a correlation between the high-velocity blue edge, \ved,  of the iron-peak Fe/Co/Ni  $H$-band emission feature and the optical light curve shape of normal, transitional and sub-luminous type Ia Supernovae (SNe~Ia).
We explain this correlation in terms of SN Ia physics.
$v_{edge}$ corresponds to the sharp transition between the complete and incomplete silicon burning regions in the ejecta. 
It measures the point in velocity space where the outer \Nifs\ mass fraction,  $X_{\rm{Ni}}$, falls to the order of 0.03-0.10. 
For a given \Nifs\ mass, $M(^{56}Ni)$, \ved\ is sensitive to the specific kinetic energy \KE($M(^{56}Ni)/M_{WD}$) of the corresponding region.
Combining $v_{edge}$ with light curve parameters (i.e., \sBV, $\Delta m_{15,s}$ in $B$ and $V$) allows us to distinguish between explosion scenarios. 
 The correlation between $v_{edge}$ and light-curve shape is consistent with  explosion models near the Chandrasekhar limit.
 However, the available sub-$M_{Ch}$ WD explosion model based on SN\,1999by exhibits velocities which are too large to explain the observations. 
 Finally, the sub-luminous  SN\,2015bo
exhibits  signatures of a dynamical merger of two WDs  demonstrating
diversity among explosion scenarios at the faint end of the SNe~Ia population.
\end{abstract}


\section{Introduction} 
\label{sect:intro}

Significant evidence supports the idea that
type Ia supernovae (SNe~Ia) result from 
%Type Ia supernovae (SNe Ia) are 
the thermonuclear explosions of at least one carbon-oxygen (C-O) white dwarf (WD) in a binary system. 
There are two main progenitor channels which have been hypothesised to produce these
cosmic explosions. 
%luminous displays. 
These are the single degenerate scenario (SDS), and the  double degenerate scenario (DDS). In the SDS a WD accretes material from a non-degenerate companion star such as a H/He or red giant star \citep{Whelan73,Livne90,Woosley94,Nomoto97}, whereas in the DDS the system consists of two WDs.

Within each progenitor scenario, multiple explosion mechanisms have been explored. 
One method for exploding a SN~Ia is with heat released during the dynamical merging
of two WDs  \citep[\eg][]{Dan14,Dan15}.

Alternatively, when a WD approaches the Chandrasekhar mass ($M_{Ch}$) the explosion can be triggered by compressional heating in the center.  This possibility may occur in both the SDS \citep{hk96,Nomoto97}, as well as in the DDS scenario where  a tidally disrupted WD accretes onto the primary WD on secular time scales, which are much longer than the hydro-dynamical time scales \citep{Piersanti03}.
In a $M_{Ch}$ explosion the nuclear burning flame front  begins as a sub-sonic  deflagration wave and then at a particular transition density, $\rho_{tr}$, it evolves into a supersonic  detonation wave. This is known as a delayed detonation  (DDT) model and has been shown to provide a good match to spectra and light curves of SNe~Ia \citep{Khokhlov91,Yamaoka92,hk96,Gamezo05,Poludnenko11,Blondin11,Blondin15,Hoeflich17,Ashall18}. 
For DDT explosions 
the luminosity of the SN is correlated with the amount of burning during the deflagration
phase  \citep{Gamezo05,Ropke07,Jordan08},  which in spherical models is parameterized by $\rho_{tr}$, 
where  less luminous objects have a lower $\rho_{tr}$ value \citep{hk96,Hoeflich17}.
Effectively, a low  $\rho_{tr}$ produces more intermediate mass elements (IME)  at the expense of \Nifs, and for less luminous objects the remaining \Nifs\ is located at lower velocities. 

Another possibility is the explosion of a sub-$M_{Ch}$ WD in the so-called edge-lit, 
or double detonation scenario.
Here, He accreted from a companion star onto the surface of the WD detonates, which drives a
shock wave inward igniting the centre of the WD,
while the outer layer is consumed by the initial detonation
\citep{Woosley94,Livne95,hk96,nughydro97,Fink2007,Pakmor12,Shen14}. These models have recently made a revival as it has been shown that by mixing the outer He layer with a small amount of C alters the burning network from slow triple-$\alpha$ to the fast $^{12}C(\alpha,\gamma)^{16}O$ channel \citep{Shen14}. 
This reduces the mass of the He shell required to form a sustained nuclear detonation by an order of magnitude compared to previous work \citep[\eg][]{Woosley94,Livne95}.
The resulting density structures are close to spherical with small polarization \citep{Bulla16}, and to first order, the outer layers hardly affect the light curves (LCs) beyond $\approx 1$ week after the explosion \citep{Polin18}. However,  it is not clear whether the reduced He mass can trigger a detonation in the C/O layers of the WD.  
In sub-$M_{Ch}$ models the luminosity of the SN is correlated with the ejecta mass of the explosion, where less luminous objects have a smaller WD mass (M$_{WD}$); \citep{Sim10,Blondin17}.

The luminosity of a SN\,Ia is dependent on the amount of \Nifs\ synthesized in the explosion. 
More luminous SNe produce larger amounts of \Nifs\ \citep[\eg][]{Arnett82,stritzinger06,Mazzali07,Childress15}. 
Furthermore, events which are more luminous also have broader light curves, which 
is the underlying basis for the luminosity-width-relation (LWR) \citep{Phillips93,Phillips99}.
The LWR can be understood in terms of opacities, where brighter objects have more  \Nifs, produce more heating, are dominated by doubly ionized species, and therefore have slowly evolving light curves.
Whereas, the faintest SNe~Ia have less  \Nifs, less heating,   are dominated by  singly ionized species, and have faster light curves.  \citep[][]{Nugent97,Umeda99,Kasen09,Hoeflich17}.

At the faint end of the LWR, there are sub-luminous SNe~Ia \citep[1991bg-like;][]{Filippenko92,Leibundgut93,Turatto96}.
The literature contains a number of different scenarios accounting for the origins of sub-luminous  SNe~Ia. These extend from the dynamical merger of two WDs\footnote{For normal SNe~Ia this is currently out of favour as it is inconsistent with low  polarization observations \citep{Patat12,Bulla16a}. } \citep{Garcia-Berro17}, to sub-$M_{Ch}$ explosions \citep{Scalzo14,Blondin17,Blondin18}, and to  $M_{Ch}$ DDT explosions \citep{Hoeflich02,Hoeflich17}. 
A key to understanding SNe~Ia is to address the on-going question of whether normal, transitional\footnote{Transitional SNe Ia are thought to be the link between the normal and sub-luminous populations (e.g., see \citealt{Hsiao15,Ashall16a,Ashall16b,Gall18}).} and sub-luminous SNe~Ia are separate groups, form a continuum, or are a mixture of diverse scenarios. This work aims to address this question.

The ongoing discussion about explosion 
scenarios in SNe Ia physics may, at least in part,
be attributed to different assumptions in the 
modeling process.
These assumptions include LTE population levels \citep[\eg][]{Goldstein18},
a small atomic network of isotopes  \citep[\eg][]{Polin18},
and that  $M_{Ch}$
explosions do not have a central core of electron capture elements. 
On the other hand, we know non-LTE effects are important,
and, even full non-LTE simulations result in different conclusions on 
the nature of sub-luminous SNe Ia,
see \citet{Hoeflich02,Blondin17,Hoeflich17,Blondin18}.

NIR spectroscopy offers a promising way to investigate the physics of SNe~Ia.
In the $H$ band, between maximum light and +10d,  an emission component is formed
by blends of a large number of emission lines above the photosphere \citep{Wheeler98,Hoeflich02,Hsiao09,Hsiao13}.
This $H$-band feature consists of a multiplet of many allowed \FeII/\CoII/\NiII\ lines formed within the \Nifs-rich layers and is thus correlated  with the luminosity of the SN \citep{Hsiao13,Ashall19}.  

One of the main objectives of the \textit{Carnegie Supernova Project II} (CSP-II; \citealt{Phillips18}) was to obtain a large sample of NIR spectra of SNe~Ia \citep{Hsiao18}. Using this data, \citet{Ashall19} found a correlation between the outer blue-edge velocity, \ved, 
of the $H$-band break region and the optical light-curve shape for normal, transitional and sub-luminous SNE~Ia. \footnote{We note that this work does not include analysis of other SNe~Ia sub-types objects such as SN\,1991T or SNe Iax.}
Here, we explain this correlation in terms of SNe~Ia physics and models. 
We compare the data to both spherical sub-$M_{Ch}$ and $M_{Ch}$ explosion models with non-LTE light curves and spectra published in the literature. 
We use 1D calculations as they artificially suppress mixing and produce a chemically layered structure. This is 
in line with both observations \citep{Fesen07,Maeda10,Diamond18,Dhawan18} and  inferred abundance stratification results \citep[\eg][]{Stehle05,Tanaka11,Ashall16b}. This suppressed mixing may  be due to high magnetic fields \citep{Hristov18}. 

In this work, we show how \ved\ varies among different explosion models,
even for SNe with similar \Nifs\ masses.
Therefore  \ved\ provides new information
beyond the total amount of \Nifs\  synthesized in the explosion, which has been classically used to analyze SNe~Ia \citep{stritzinger06,Childress15,Dhawan17,Scalzo19}.
A second goal of this paper is to put \ved\  into context with the classical characterization using
 SNe~Ia light curve shape and absolute magnitude, and to show an  example of
how this combined information can be used to distinguish between explosion models.
\\
\section{Measurements of Spectra and Light Curve Parameters}
\label{sect:2}

\begin{figure}
\centering
\includegraphics[width=\columnwidth]{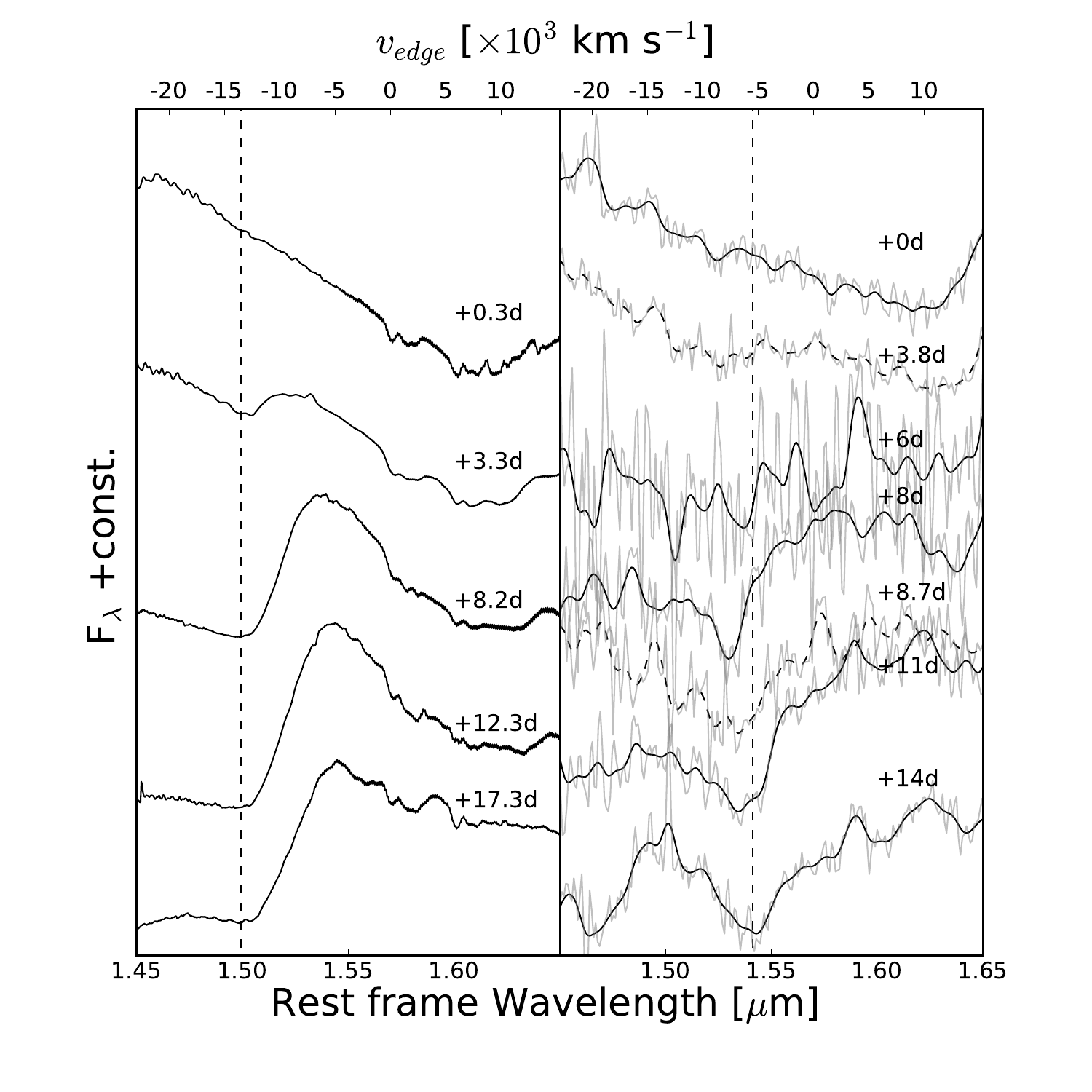}
\caption{A comparison of the $H$-band region of a normal SN Ia (left) and a sub-luminous SN Ia (right). \textit{Left:} A time series of spectra of SN\,2011fe  \citep{Hsiao13}. The vertical dashed line is the blue-edge of the $+$12.3\,d spectra ($-$13,500~\kms).  \textit{Right:} A combined time series of spectra 
for SN\,1999by (solid; \citealt{Hoeflich02}) and ASASSN-15ga (dashed). The spectra were  Gaussian smoothed (2$\sigma$), and the underlying un-smoothed spectra are plotted in light grey. The vertical dashed line is the blue-edge of the +11\,d spectra at $-$5,500~\kms. Times are given relative to rest frame $B$-band maximum.}
\label{fig:11fevs99by}
\end{figure}

In this work, the $H$-band break is represented by the Doppler shift, \ved,  of the bluest component of the $H$-band multiplet at 1.57 \mic\ (Fig. \ref{fig:11fevs99by}). 
In a SN~Ia explosion, the photosphere recedes through the IME layers before it reaches the \Nifs\ region. Once this \Nifs-rich region is exposed, the Fe/Co/Ni emission in the $H$-band
begins to emerge.
\citet{Ashall19} found that at about +10 days relative to $B$-band maximum was a good time to measure \ved, because during this phase the spectra are dominated by single ionized iron group elements,  and by then the feature has  emerged in all SNe~Ia. 
However, note that at earlier times, \ved\ may be affected by lines of IME.  Whereas, significantly beyond +10d  the emission from forbidden lines starts to dominate.
In fact, the ideal time to measure \ved\ may be right after the H-band feature can be clearly distinguished from other blends. However, this would require daily observations which are  not currently available.

We measure \ved\ by the method outlined in  \citet{Ashall19} in both the observed and theoretical spectra. Briefly, \ved\ is measured by fitting a Gaussian profile to the minimum of the blue-edge of the $H$-band break. The fit is produced over a fixed wavelength range and iterated to find convergence.  
The model spectra are
based on detailed non-LTE radiation transport explosions for normal, transitional and sub-luminous SNe~Ia \citep{Hoeflich02,Hoeflich17,Blondin18}.
For all models close to $M_{Ch}$, \ved\ corresponds to 
the region where the \Nifs\ mass fraction, $X_{\rm{Ni}}$,
is approximately 0.02-0.03.
For the sub-luminous sub-$M_{Ch}$ model with low mass ($M_{WD}$=0.9\Msun) \citep{Blondin18} the corresponding abundance is slightly higher ($X_{\rm{Ni}}$ is $\approx $ 0.10)\footnote{We note that \citet{Blondin18} use  incomplete nuclear reaction networks, and complete networks predict a larger \Nifs\ production \citep{Shen18b}. However, the velocity where $X_{\rm{Ni}}\approx $ 0.10 in the complete networks is consistent with that of \citet{Blondin18} within 500\kms.}. 
This is  because the lower mass produces a smaller optical depth.
 For all models, regardless of $M_{WD}$, the value of $X_{\rm{Ni}}$ is small. 
Therefore, we use the H-band break as an indicator for the outer edge of the \Nifs\ region.  
For a discussion of the uncertainties in the fluxes see the references for each of the models.

\begin{figure}
\centering
\includegraphics[width=\columnwidth]{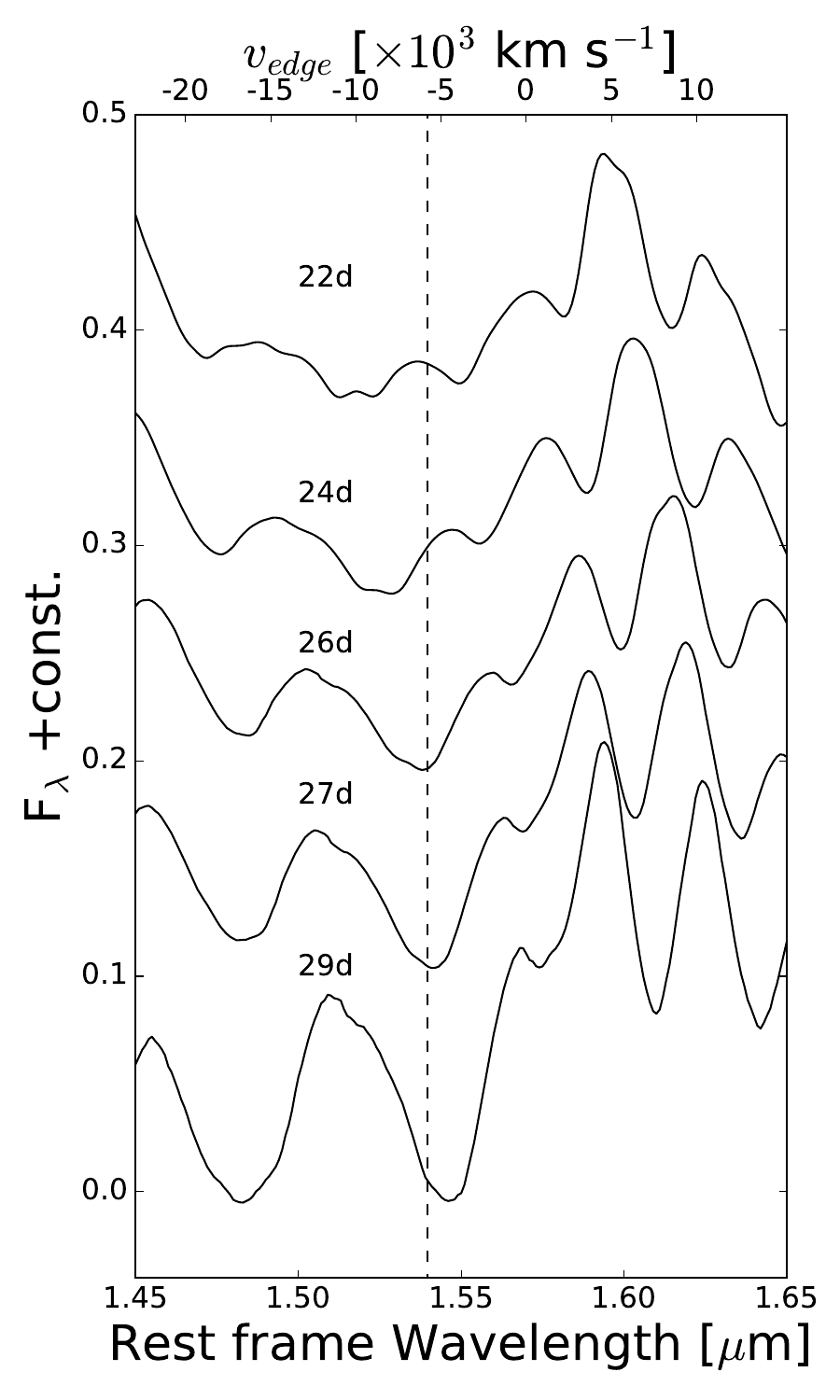}
\caption{A time series of model spectra of a sub-luminous SN Ia from \citet{Hoeflich02}. It can be seen that at the earlier epochs the value of \ved\ could be affected by blends from IMEs. Times are  given relative to the explosion. The rise time of the SN is \ab 14.5$\pm$0.5day. The vertical dashed line is value of \ved\ ($-$5800\kms) at 26 days past explosion, or +11.5$\pm$0.5\,d from maximum.}
\label{fig:model}
\end{figure}

\begin{figure*}
\centering
\includegraphics[width=\textwidth]{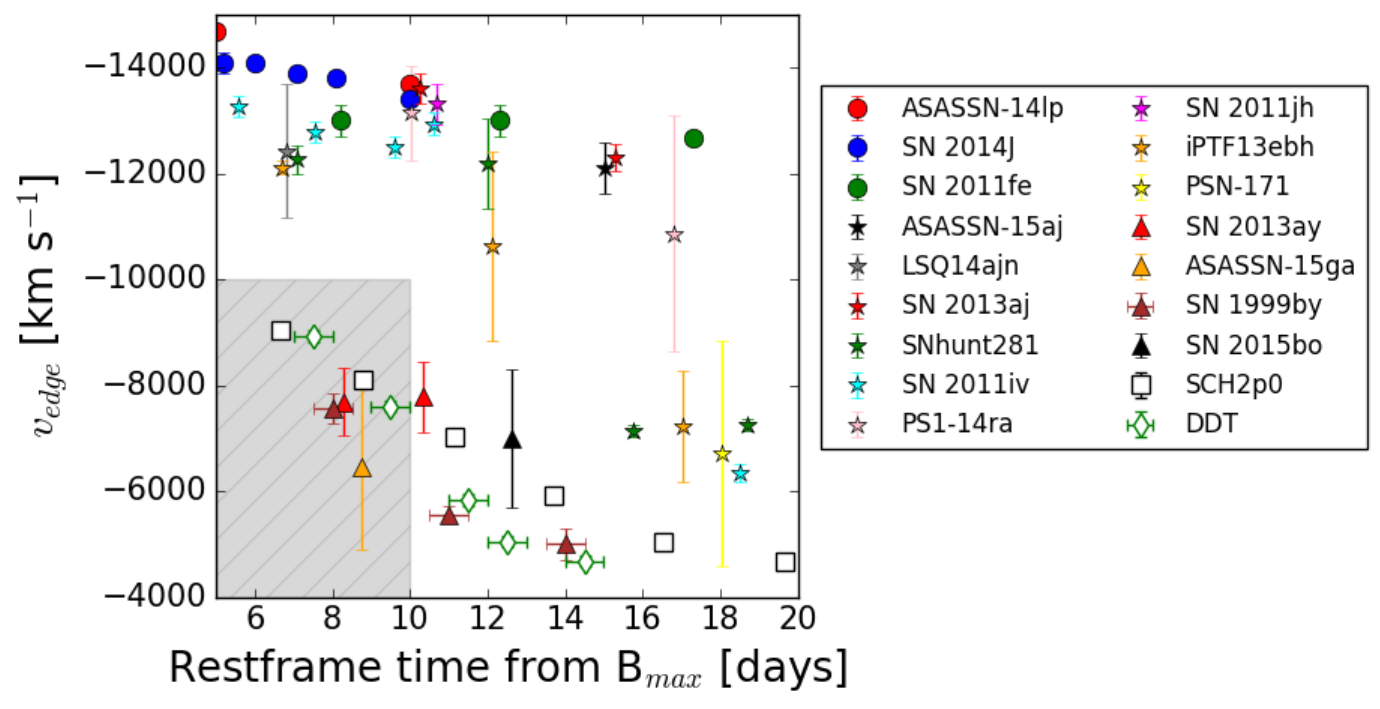}
\caption{\ved\ as a function of time from \citet{Ashall19}. For comparison, the $M_{Ch}$ (open green diamonds) and sub-$M_{Ch}$ (open black  squares) models of the sub-luminous SN\,1999by from \citet{Hoeflich02} and \citet{Blondin18} have been plotted. The grey shaded region is the area where the spectra may be affected by line blends and may not be suitable to be used for measuring the \Nifs\ abundance, see section \ref{sect:2} for details.  For the sub-luminous SNe Ia, in the epochs where the measurement of \ved\ is reliable, it is apparent that the sub-$M_{Ch}$ model has values larger than the observations. Whereas the $M_{Ch}$ model arein agreement with the observations. Normal SNe~Ia are marked by solid circle symbols, transitional SNe~Ia are marked by star circle symbols, and sub-luminous SNe~Ia are marked by solid triangle symbols.}
\label{fig:timeee}
\end{figure*}

In Fig. \ref{fig:11fevs99by}, we present observations 
of the spectral evolution for a normal-bright and a sub-luminous SN~Ia. 
For bright SNe~Ia, the iron-peak feature emerges at about +3 days past maximum light when the overall  NIR continuum
is dominated by Thomson scattering and, thus, will have a very little line blending beyond the blue edge. For the sub-luminous example, blends of other transitions are obvious in both the observations and the theoretical model,
because they have a lower ionization state and less Thomson scattering (Fig. \ref{fig:model}). These blends can be seen until about +10 days, and  can be attributed to lines of singly-ionized iron-group elements, \CaII\ and \CaI\ (1.46, 1.51 \mic) and neutral IME elements, namely \SiI\ (1.52 \mic), \SI\ (1.46, 1.54 and 1.52, 1.57 \mic) \citep{Hoeflich02}. Therefore, for sub-luminous SNe~Ia spectra should be inspected to make sure there are no blends of IMEs\footnote{We note that as these IME lines are very sensitive to the ionization state,  whether epochs earlyer than +10\,d can be used to measure \ved\ can be decided by the spectra of each specific SN.}. 
 However, even for the least luminous objects at +10\,d after maximum it is predicted that there will be virtually no contamination by IMEs. Hence, this is a good time to measure \ved.
The effect of blends as a function of time can be seen in Figs. \ref{fig:11fevs99by}, \ref{fig:model}, and \ref{fig:timeee}.  For sub-luminous SNe~Ia,  these blends mimic a rapid drop in the evolution of \ved\,  before +10\,d as can be seen in the velocity evolution of SN\,1999by (see Fig. \ref{fig:timeee}).
 For transitional SNe~Ia, such as SN\,2011iv, SNhunt281, and iPTF13ebh, the observed rapid drop in \ved\ as a function of time can be understood as follows: \ved\ is larger and is thus at lower densities compared to sub-luminous supernovae. These densities  are close to the critical density ($\leq 10^{7...8} particles/cm^3$) which marks the transition from allowed to forbidden lines. This results in a rapid drop of \ved. We note that, in normal-bright SNe,
the temperatures are higher resulting is larger collisional rates which de-populate the levels responsible for forbidden lines and delay the rapid drop in  \ved.

In this work, we use the \sBV\ parameter defined in \citet{Burns14} and 
used in \citet{Ashall19}, 
as well as $\DmsB$  and $\DmsV$ . 
$\DmsB$ and $\DmsV$ \citep{Hoeflich10,Hoeflich17}, are modified decline rate parameters produced for a time base of $t=15~d*s$, with $s$ being the stretch parameter of the light curve as defined in \cite{Goldhaber01}. 
They can be parameterized by $\DmsB={\rm \frac{\Delta m_{(15\times s)}}{s}}$. 
Normal SNe Ia have an $\DmsB$ and $\DmsV$ of \ab1.2\,mag and 0.7\,mag, respectively. 
We chose to use $\DmsB$ and $\DmsV$ as they can be theoretically interpreted as 
a measurement of diffusion time scales \citep{Hoeflich17}, 
whereas the theoretical interpretation of $s$ or \sBV\ is not straight forward but a combination of effects.

\section{Brief Description of the Explosion Scenarios and their relation to light curve parameters}
\label{sect:models}

Normal, transitional, and sub-luminous SNe~Ia can be thought to be in three different regimes. In the case of normal SNe~Ia the opacity of the ejecta is dominated by doubly ionized species throughout most of the envelope, leading to a slowly declining post-maximum light curve. Transitional SNe~Ia are
located in the area of rapidly decreasing luminosity in the LWR \citep{Hoeflich02,Hoeflich17,Ashall18}).
This is the region where a SN~Ia enters the regime of a quickly decreasing opacity in the envelope soon after maximum, due to  an earlier onset of the recombination front that results in 
a faster release of stored energy. Whereas, sub-luminous SNe Ia are dominated by singly ionized species, which leads to a fast declining post-maximum light curve.

Within spherical DDT models,
for the brightest SNe~Ia most of the \Nifs\ is made in the detonation phase. 
Furthermore, in these models the \Nifs\ production is an almost
smooth function of $\rho_{tr}$, but its relative change is quick between models.
This is reflected by a rapidly decreasing luminosity over the transitional regime, which, to first order, is $\propto M($\Nifs$)$. 
In other words, the \Nifs\ production during the detonation phase changes from \ab0.3 to 0 $M_\odot$, as $\rho_{tr}$ decreases from \ab 1.8$\times$10$^{7}$g\,cm$^{-3}$ to 0.8$\times$10$^{7}$g\,cm$^{-3}$ over the transitional regime towards sub-luminous SNe Ia.
For sub-luminous SNe Ia  almost all of the \Nifs\ is produced during the deflagration burning phase, which happens over a
\ab$0.25\,M_\odot$ range. 
This leads to a close to constant luminosity for sub-luminous SNe~Ia and a  \Nifs\ production of 0.1-0.2~\Msun\ \citep[\eg][]{Stritzinger06b},  which is dependent on the rate of electron capture \citep{Hoeflich17,Gall18}.

In the sub-$M_{Ch}$ models considered here a central explosion is triggered in a static WD \citep{Blondin18}.
These types of models are used as a proxy for a helium detonation, as well as the dynamically driven double-degenerate double-detonation scenario \citep{Shen18a}.
In these classes of models the mass of the WD is correlated to the luminosity and the light-curve shape. Lower mass WDs produce less effective burning, 
less nuclear statistical equilibrium (NSE) elements, less \Nifs, less heating, lower opacities, and a faster light curves.

\begin{figure*}
\centering
\includegraphics[width=\textwidth]{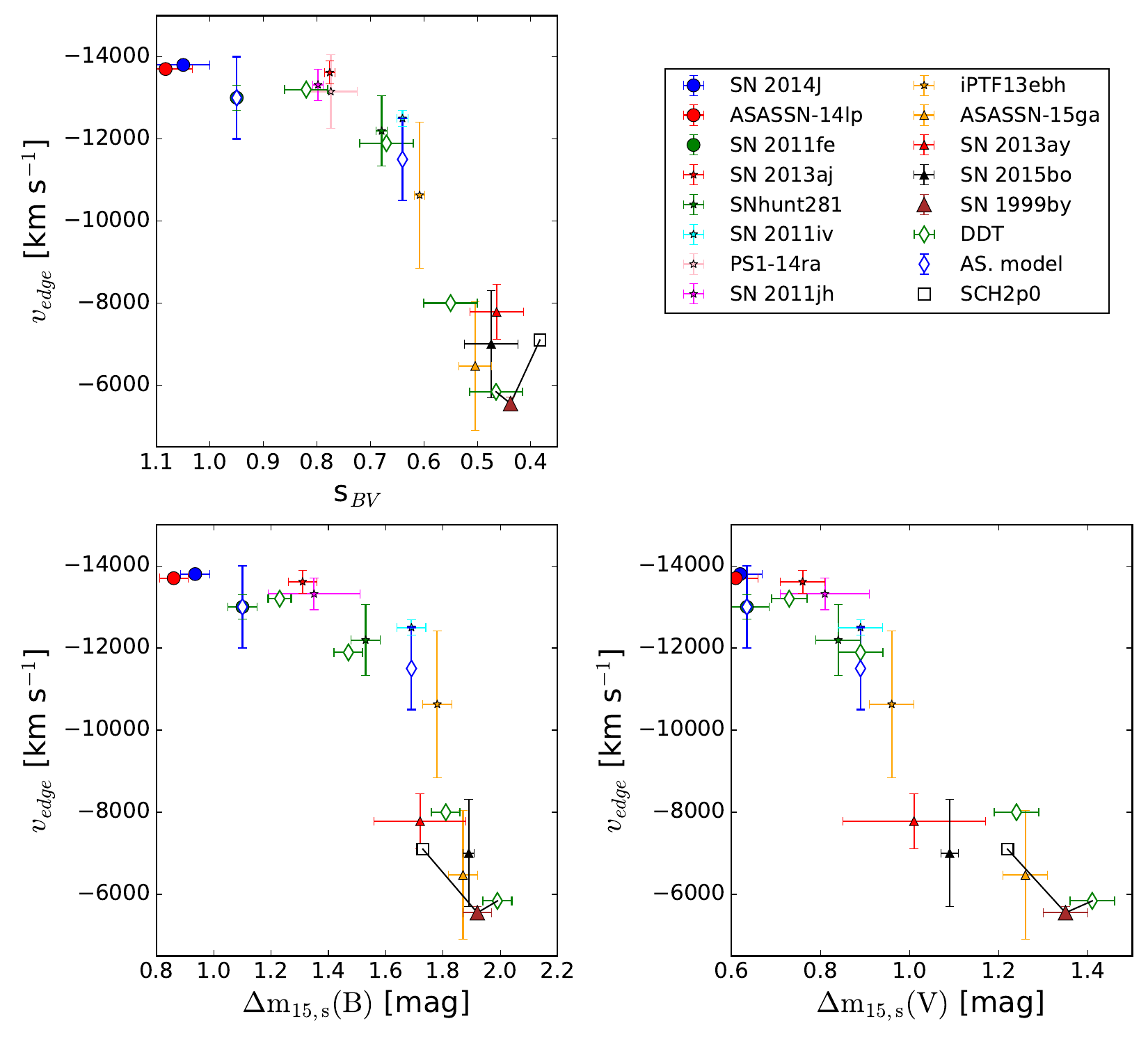}
\caption{  
\textit{Upper left:} The iron-peak outer velocity at  +10$\pm$3\,d as a function of \sBV, the open symbols are models and filled in symbols are observations. The open green diamonds are  non-LTE $M_{Ch}$ DDT models from \citet{Hoeflich17}, the open blue diamonds are the results from abundance stratification models for SN\,2011fe and SN\,2011iv from \citet{Mazzali14} and \citet{Ashall18}, respectively, and the open black square is the SCH2p0 (0.9\Msun) model from \citet{Blondin18}. The solid black lines links SN\,1999by with its  sub-$M_{Ch}$ and  $M_{Ch}$ models. These models 
therefore have the same luminosity as SN\,1999by. 
\textit{Lower left:} The same as the top left panel but as a function of $\DmsB$.
\textit{Lower right:} The same as the top left panel but as a function of $\DmsV$. In all panels the $M_{Ch}$ models follow the observations, but the sub-$M_{Ch}$ model has a velocity which is \ab1,500\kms\ larger than SN\,1999by.   Whereas the $M_{Ch}$ model follows the observations. Normal SNe~Ia are marked by solid circle symbols, transitional SNe~Ia are marked by star circle symbols, and sub-luminous SNe~Ia are marked by solid triangle symbols.}
\label{fig:veltime}
\end{figure*}

\begin{figure}
\centering
\includegraphics[width=\columnwidth]{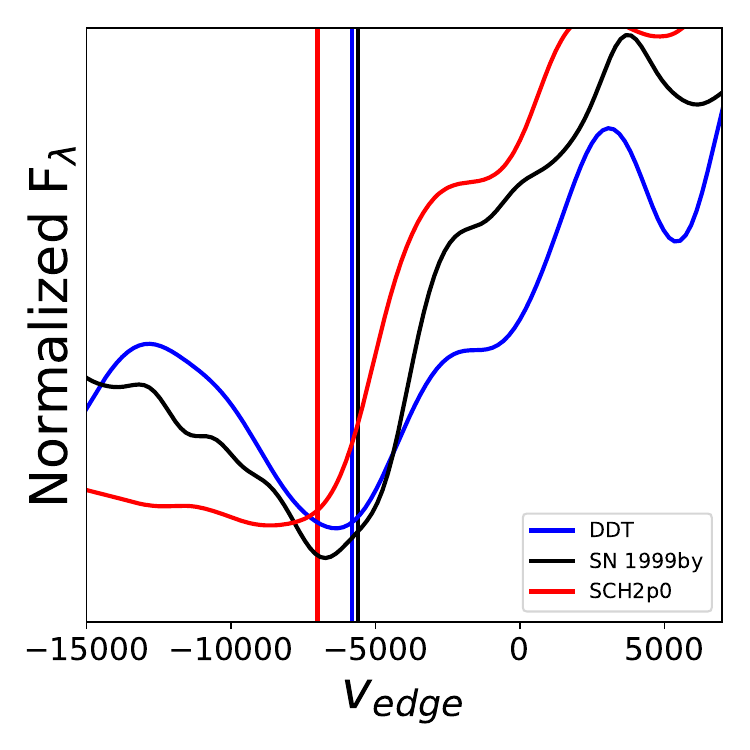}
\caption{A comparison of the spectra of SN\,1999by (black) at +11d and models of $M_{Ch}$ DDT explosion from \citet{Hoeflich02} (blue) and sub-$M_{Ch}$ SCH2p0 explosion from \citet{Blondin18} (red). The spectra have been Gaussian smoothed with a sigma of 3. The vertical solid lines are located at the measured velocity of the minima. The velocity of the sub-$M_{Ch}$ model (\ved=$-$7000$\pm$200\kms) is larger than both the observations (\ved=$-$5500$\pm$200\kms) and the $M_{Ch}$ DDT model (\ved=$-$5800$\pm$200\kms). We note that vedge is not at the exact minimum of the feature because a continuum is subtracted in the fitting procedure, see \citet{Ashall19}. This  makes the value of \ved\ in the sub-Ch model closer to observations than the plot shows.}
\label{fig:Hthing}
\end{figure}
\section{The relation of \ved\ and light curve parameters}
\subsection{\ved\ versus \sBV}
\label{sect:vedsbv}

Here, we  discuss \ved\ as a function of light-curve color-stretch parameter, \sBV.
As shown in \citet{Ashall19} and as illustrated 
in the  upper left panel of in Fig.~\ref{fig:veltime},
SNe with a larger \sBV\ are found to exhibit higher values of \ved. 

A measure of \ved\ serves as a powerful diagnostic tool because it is robust due to Doppler shifts that are well measured by spectra at the onset of the emergence  of the $H$-band feature. 
For a comparison between data and theory, in Fig.~\ref{fig:veltime} we  plot
results obtained from non-LTE, spherical $M_{Ch}$ DDT models \citep{Hoeflich17}, as the open green diamonds.
Note that in this figure the values of \ved\ are obtained from four different DDT models (Model 23, 20, 12, and 8) of \citet{Hoeflich17}, whereas in Figure \ref{fig:timeee}, one sub-luminous DDT model and one sub-luminous sub-$M_{Ch}$ model are plotted as a function of time.

A comparison with the abundance structures of the DDT models shows that \ved\  measures the point in 
velocity space where $X_{Ni}$ falls to the order of 0.02 to 0.03, for the entire range from normal-bright to sub-luminous SNe~Ia.

For  burning to NSE, the density must be larger than $\approx 2 \times 10^7 g/cm^3$.
In the DDT scenario,
for normal-bright SNe~Ia, \ved\ corresponds to a region where the 2 to 3 \% of \Nifs\ required to form the emission feature is located between $\approx -13,000$ to $ -10,000$~\kms. 
Because there is little mass involved in these layers, the change in the total amount of \Nifs\ produced over this \ved\ range is small and, consequently, the luminosity difference between normal SNe~Ia is little.

There is a fast drop in \ved\  over an \sBV\ range of \ab0.6 to 0.45,  
the mean $v_{edge}$ drops from \ab$-$11,500~\kms\ to \ab$-$5,500~\kms.
This is because the change in mass per unit velocity rapidly decreases
with increasing velocity.  
From DDT models, the bright-transitional SNe~Ia are characterized by an NSE production that is dominated by detonation burning, whereas the  less luminous transitional supernovae are dominated by deflagration burning.
Because the densities are still low, 
a large relative change in the total NSE mass coincides with a large shift in \ved. 
Therefore, for transitional SNe~Ia a uniform change in $\rho_{tr}$ produces a significant drop in \ved. 
This drop in \ved\  happens in the same regime as the luminosity drop in the LWR. 
However, these processes are not causally linked. 
Hence, \ved\ is a complementary measurement of the SNe~Ia physics and scenario, and it is not a proxy for the light-curve shape. 
This is because the diffusion time scales depend on the \Nifs\ distribution, \Nifs\ mass, and ejecta mass., whereas, \ved\  only depends on the outer edge of the  \Nifs\ region, and measures the specific \KE.
Finally, for sub-luminous SNe~Ia, most of the NSE burning takes place during the deflagration phase in the inner, high-density regions, 
which produces a minimum value of  \ved\ at \ab$-$5,500~\kms.

Two of the SN (SN\,2011fe and SN\,2011iv) in our work have been analyzed using the abundance stratification technique \citep{Mazzali14,Ashall18}. This method reverse engineers the abundance structure in the ejecta, using a comparison between observed and synthetic spectra \citep[\eg][]{Stehle05,Ashall14,Ashall16b}. 
We also plot the velocity of \Nifs\ at 3\% in abundance from these models (see the blue open diamonds in Fig.~\ref{fig:veltime}). There is a good agreement, within \ab1000\,\kms\, with the abundance stratification results, non-LTE DDT models, and observations. It should be noted that by definition the abundance stratification models have the same luminosity as their corresponding SNe.

\subsection{SN\,1999by}
\label{sect:99by}
One supernova from the sample, SN\,1999by a sub-luminous SN, has been modeled by both
$M_{Ch}$ \citep{Hoeflich02} and sub-$M_{Ch}$ \citep{Blondin18}  explosions, which makes it a perfect test case for this work.  
Therefore, included in Fig.~\ref{fig:veltime} is  the sub-$M_{Ch}$ model (SCH2p0) for the sub-luminous SN\,1999by from \citet{Blondin18} 
(open black square).
The SCH2p0 ($M_{WD}$=0.9\Msun) sub-$M_{Ch}$ model has a velocity of  $-$7,000~\kms, which is much larger, by 1,500 \kms, than the observations.
In this model \ved\ corresponds to 
the region where the $X_{\rm{Ni}}$ is 0.10.  This is due to a 
combination of geometric dilution and 
the lower mass meaning that  a larger mass fraction is
required to form the emission feature.

 Fig. \ref{fig:Hthing} presents the spectra of SN\,1999by (black)  as well as the $M_{Ch}$ DDT model (blue) from \citet{Hoeflich02},
and the sub-$M_{Ch}$ model, SCH2p0, (red), from \citet{Blondin18}, all of which are at \ab+11d relative to $B$-band maximum. Note that the models from \citet{Hoeflich02} and  \citet{Blondin18} have similar $B$-band rise times of 14.5\,d and 14.6\,d,  respectively. It is clear that the  
$M_{Ch}$ DDT model produces the correct velocity compared to the observations, whereas the sub-$M_{Ch}$ model has a value of \ved\ which is too large. 

As  shown by both the $M_{Ch}$ and sub-$M_{Ch}$ models,  \ved\ is determined by the interplay between the mass of \Nifs\ 
and the mass of IME formed in the explosion,  because there is a steep drop in the abundances during the transition between 
nuclear statistical equilibrium and incomplete Si-burning.
This makes \ved\  a stable measurement regardless of the explosion scenario.
In effect, \ved\ is a measure of the specific kinetic energy,  \KE\  $(M($\Nifs$)/M_{WD}$), of the region. 
For a given M(\Nifs), models with lower values of $M_{WD}$ will result in systematically higher values of \ved.

\subsection{\ved\ \&\ $\DmsB$, $\DmsV$}
\label{sect:dm15s}

\sBV\ measures the timing of the turnover in the color curves of SNe Ia, but as we will see below, may mask diversity. Therefore, we  use the light curve parameters $\DmsB$ 
and $\DmsV$  to analyze the correlation.
The bottom panels in  Fig.~\ref{fig:veltime} present \ved\ as a function of
$\DmsB$ and $\DmsV$. 
Once again the DDT models fit the data well, where as the sub-$M_{Ch}$ model is not close to SN\,1999by. 

Nine out of ten SNe~Ia in the plot  are  consistent with $M_{Ch}$ DDT models.
However,  SN\,2015bo, which has a high-cadence  light curve, with pre-maximum coverage, and will be the subject of a future individual analysis,  has a large $\DmsB$=1.89\,mag but a small $\DmsV=1.09$\,mag. 
The combined values of light curve shape and \ved\ 
for SN\,2015bo are inconsistent with both the DDT and sub-$M_{Ch}$ models.
For a \ved=$-$7000\kms\ DDT models predict a 
$\DmsB=\sim1.85$\,mag and $\DmsV=\sim1.32$\,mag, and 
sub-$M_{Ch}$ models predict 
$\DmsB=\sim1.73$\,mag and $\DmsV=\sim1.22$\,mag.
It is difficult to reconcile this deviation from the data and models by varying the physics of sub-$M_{Ch}$ or $M_{Ch}$ DDT explosions.

Dynamical merger models are characterized by red colors at maximum light, a slowly evolving
 $V$-band light curve, a fast declining $B$-band light curve, and have \Nifs\ located at low velocities \citep{hk96,Garcia-Berro17}.
 These properties make dynamical mergers a viable scenario for SN\,2015bo, and demonstrate that there may be different explosion mechanisms and progenitor scenarios present within the sub-luminous SNe Ia population.

We note that, on its own, \ved\ may not be able
to distinguish between different models. 
The method here also requires accurate high-cadence light curves. 
For example, in the case of SN~2013ay where the photometric coverage begins at \ab+10\,d, 
we cannot rule out either the sub-$M_{Ch}$ or $M_{Ch}$ mass models,
despite having high signal-to-noise NIR spectra.

The diversity among SNe~Ia is apparent when we combine
$\Delta m_{15,s}$ and \ved, and extra information may be obtained if 
 the absolute luminosity ($M_{V,B}$) of the supernova is also utilized. 
Therefore, for further insights, we suggest that all three parameters are used.
In the one supernova, SN\,1999by, where we can accurately determine the absolute magnitude from observations, it is clear that the $M_{Ch}$ model is favored.
Combining absolute magnitude, with \ved\ and light-curve shape will allow for future work to more accurately discriminate between explosion scenarios and models.
For example, a transitional $M_{Ch}$ explosion  might have the same value of \ved\ as a sub-luminous sub-$M_{Ch}$ explosion, and both objects may have  similar light-curve shapes,  but different \Nifs\ masses and absolute magnitudes. Therefore, combining all parameters will provide additional information.

\section{Conclusion}
\label{sect:conclusion}

Using the correlation found between light-curve shape and \ved\ in  \citet{Ashall19}, 
we have demonstrated that
\ved\ is a new comprehensive way to measure the outer edge of the \Nifs\ region in SNe Ia ejecta. 
Brighter SNe~Ia have \Nifs\ located at higher velocities than sub-luminous SNe Ia (see Fig.~\ref{fig:veltime}). 
This  is consistent with previous results  obtained by nebular phase spectral modelling  \citep{Mazzali98,Boty17}. 
Using SN\,1999by as an example, we have also demonstrated that  a combination of \ved, $\Delta m_{15,s}$, and absolute magnitude can be
a new method to probe for diversity and explosion scenarios in SNe~Ia.

\ved\ is a stable measurement which is determined by the interplay between the mass of \Nifs\ and the mass of IME formed in the explosion, and  corresponds to the sharp transition in the ejecta between complete and incomplete Si-burning regions. 

 \ved\ measures the point in velocity space where $X_{Ni}$ falls to the order of 0.03-0.10, and is dependent on
 the ejecta mass of the explosion. 
\ved\ is stable when compared to models because it measures a Doppler shift which is dependent on the underlying abundance structure. \ved\ only
depends on the presence of \FeII\ and \CoII\ at the edge of the \Nifs\ region rather than the correct absolute flux in the model.  Furthermore, 
\ved\ is an important tool for studying SNe Ia physics as it is
independent of distance.

For sub-luminous SNe Ia, at early times, 
the edge of the $H$-band break may be 
contaminated by \CaII, \CaI, \SiI, and \SI\ lines. 
Therefore, we urge that caution should be used when choosing an epoch to measure \ved, because these line blends may artificially increase
its value.
The supernova needs to be in the \Nifs-rich region, and not affected by 
line blends  before measuring \ved.
For the least luminous SNe Ia, this occurs around +10d.
Although, this value can be sensitive to the amount of mixing
of \Nifs\ in the explosion, and each time series of spectra should be examined on a case-by-case basis. 

Within the framework of DDT models, the quick drop in the \ved\ vs. \sBV\ correlation  for transitional SNe~Ia %can be understood in terms of 
is due to the rapidly changing expansion
velocity as a function of mass. 
We demonstrated that $M_{Ch}$ models may be able to reproduce the evolution of $v_{edge}$ over the entire luminosity range of SNe~Ia.
However, the value of \ved\ obtained from the sub-$M_{Ch}$ model of the sub-luminous SN\,1999by differs from the observations by \ab1,500\kms.
This strongly favors high-mass explosions for the very sub-luminous SNe Ia. \
However, we cannot conclude that all of the sub-luminous SNe are inconsistent with sub-$M_{Ch}$ models.  It is possible that  future sub-$M_{Ch}$ models will overcome their current problems.

Additional distance-independent information can be obtained using $\DmsB$ and $\DmsV$ in combination with \ved. 
This allows for the 
diversity amongst low-luminosity SNe~Ia to be probed.
We find that SN~2015bo is inconsistent with both the spherical $M_{Ch}$ DDT models and sub-$M_{Ch}$ models considered here, but may have characteristics of a dynamical merger models,
which  adds to the evidence for diversity among the  sub-luminous population.

% Though nine out of ten of our sample that have good data are consistent with DDT models,
% . It shows a fast and slow decline in the $B$ and $V$ bands, respectively, which
% is inconsistent with both the spherical $M_{Ch}$ DDT models and sub-$M_{Ch}$ models considered here.
% However, dynamical merger models show the proper characteristics as discussed in \S~\ref{sect:dm15s}. 

One of the  limitations of our study includes 
the fact that we have only used
 published non-LTE models with NIR spectra.
However, in the future, the method presented here should be applied 
to a diverse set of explosion scenarios and models, as well as observations of SNe Ia which belong to different areas of the luminosity width relation. 

Our analysis here favors high $M_{WD}$ explosions, for  all of the supernovae examined.
However, it appears that there could be multiple explosion mechanisms amongst sub-luminous  SNe Ia, and  NIR spectra can reveal this diversity. 
We are beginning to obtain a full view of the SNe Ia phenomenon ranging from early time studies \citep[\eg][]{Hosseinzadeh17,Stritzinger18}, to mid-infrared studies \citep[\eg][]{Telesco15, Hoeflich18}, and NIR studies \citep[\eg][]{Hsiao18,Ashall19}.
Simultaneously examining all of these phenomena and comparing them to modern explosion models may let us unlock the  mysteries about what SNe Ia are, which will
enable us to  improve upon the utility of these objects  as distance indicators.
Looking towards the future, we belive that \ved\ 
will help us understand the physical nature of the SNe Ia and how they explode. 
%Looking toward the future, the measurement of \ved\ is an important tool to help solve the current pressing questions of what are  SNe Ia progenitors are and how they explode. 

\section{Acknowledgements:} 
In memory of Alexei Khokhlov, a great scientist  and friend  from whom we learned so much over the years.
We want to thank many colleagues for helpful  discussions. This  work has been supported 
in part by NSF awards AST-1008343 \& AST-1613426 (PI: M.M. Phillips), AST-1613472 (PI: E.Y Hsiao), AST-1715133 (PI: P. Hoeflich), AST-1613455 (PI:N. Suntzeff), and in part by a Sapere Aude Level II grant (PI: M.D. Stritzinger) provided by the Danish National Research Foundation (DFF).
M.D. Stritzinger is funded by a research grant (13261) from the VILLUM FONDEN. EB acknowledges partial support from NASA Grant NNX16AB25G. EB and MDS thank the Aarhus University Research Fund (AUFF) for  a Sabbatical research grant. Research by DJS is supported by NSF grants AST-1821967, 1821987, 1813708 and 1813466.
NBS and KK also thank George P. and Cynthia Woods Mitchell Institute for Fundamental Physics and Astronomy for support during this research.

\clearpage
\renewcommand{\thefigure}{A\arabic{figure}}
\setcounter{figure}{0}

\appendix

\begin{figure}
\centering
\includegraphics[width=10cm]{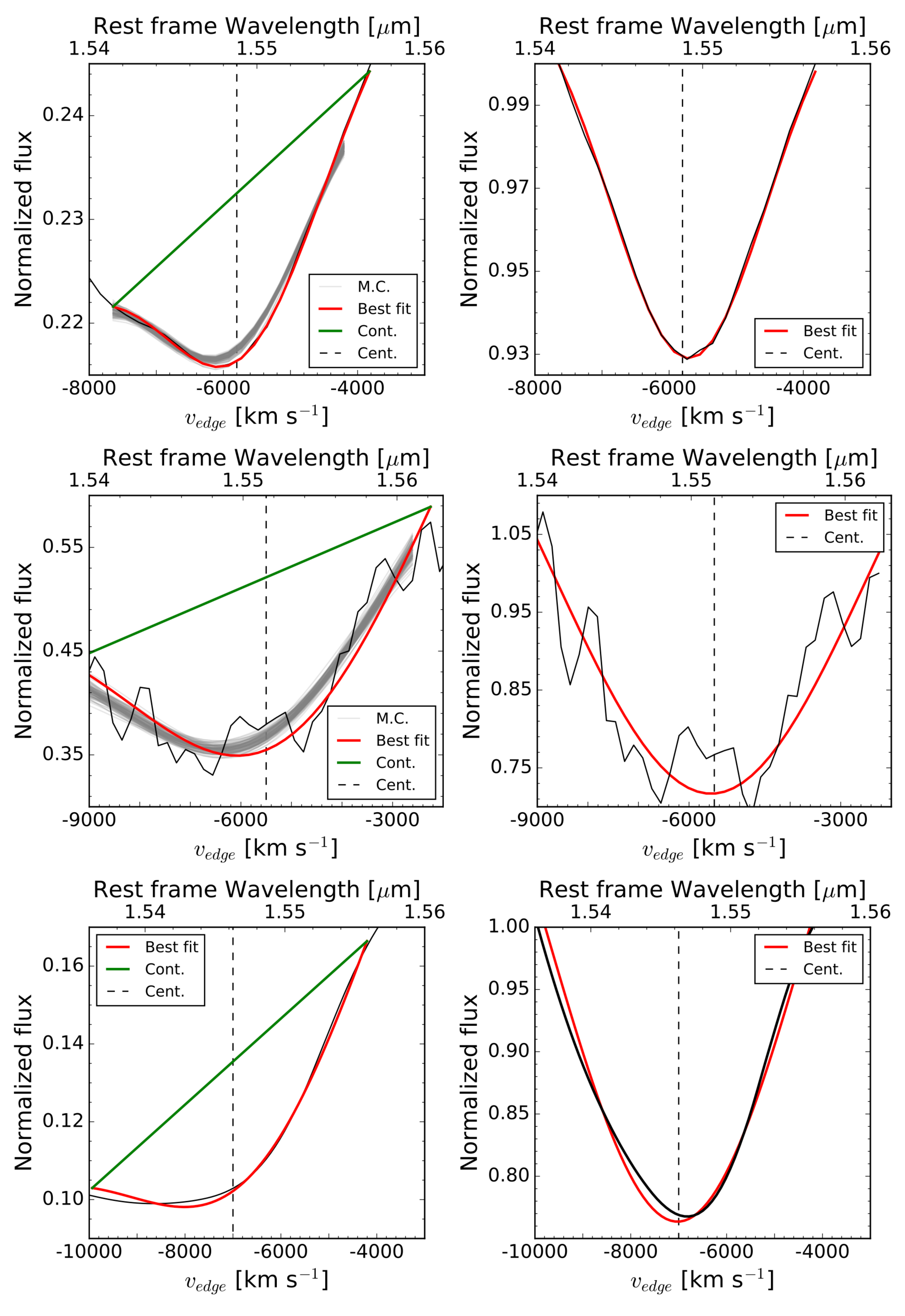}
\caption{The \ved\ fits of the sub-luminous $M_{Ch}$ model (top panels), SN\,1999by (middle panels), and the sub-luminous  sub-$M_{Ch}$  model (bottom panels). The black lines are the data or models, the red lines denote the best fits, the light grey lines correspond to the MC fits, the green lines are continuum which were subtracted, and the vertical dashed indicates the minima of the best fit and the value of \ved. }
\label{fig:apendixfit}
\end{figure}

In Figure \ref{fig:apendixfit} we present the \ved\ fits  of the $M_{Ch}$, sub-$M_{Ch}$ and data of SN\,1999by, the fitting procedure from \citet{Ashall19} was followed. The small differences between the minima of the fits and the data/models are on the order of 200\,\kms, which is well with the error bars, and therefore  does not affect the conclusions above. 

\acknowledgments

\end{document}